\documentclass[twocolumn,showpacs,superscriptaddress,preprintnumbers,amsmath,amssymb,prc]{revtex4-2}

\usepackage{graphicx}
\usepackage{dcolumn}
\usepackage{bm}
\usepackage{float}
\usepackage{color}
\usepackage{CJK}
\usepackage{ulem}
\usepackage{array}
\usepackage{makecell}
\usepackage{lineno}
\usepackage{amsmath}

\usepackage[colorlinks,linkcolor=blue,urlcolor=blue,citecolor=blue]{hyperref}

\usepackage{tikz,xcolor,hyperref}

\definecolor{lime}{HTML}{A6CE39}
\DeclareRobustCommand{\orcidicon}{
	\begin{tikzpicture}
	\draw[lime, fill=lime] (0,0) 
	circle [radius=0.16] 
	node[white] {{\fontfamily{qag}\selectfont \tiny ID}};
	\draw[white, fill=white] (-0.0625,0.095) 
	circle [radius=0.007];
	\end{tikzpicture}
	\hspace{-2mm}
}
\foreach \x in {A, ..., Z}{%
	\expandafter\xdef\csname orcid\x\endcsname{\noexpand\href{https://orcid.org/\csname orcidauthor\x\endcsname}{\noexpand\orcidicon}}
}
\foreach \x in {A, ..., Z}{%
	\expandafter\xdef\csname orcid\x\endcsname{\noexpand\href{https://orcid.org/\csname orcidauthor\x\endcsname}{\noexpand\orcidicon}}
}

\usepackage{graphicx}
\usepackage{dcolumn}
\usepackage{bm}
\usepackage{upgreek}
\usepackage{multirow}
\usepackage{makecell}
\usepackage{tabularx}
\usepackage{booktabs}
\usepackage{color}
\usepackage{longtable} 

\begin{document}

\preprint{APS/123-QED}

\title{
Precision extraction of the deuteron electric polarizability via the Baldin sum rule with full low-energy coverage}
\thanks{A footnote to the article title}%

\author{Zi-Rui Hao\orcidC{}}
 \affiliation{Shanghai Advanced Research Institute, Chinese Academy of Sciences, Shanghai, 201210}
 
\author{Gong-Tao Fan\orcidD{}}%
 \email{fangt@sari.ac.cn}
 \affiliation{Shanghai Advanced Research Institute, Chinese Academy of Sciences, Shanghai, 201210}
 \affiliation{Shanghai Institute of Applied Physics, Chinese Academy of Sciences, Shanghai, 201800}
 \affiliation{University of Chinese Academy of Science, Beijing, 101408}
 \author{Qian-Kun Sun\orcidE{}}
\affiliation{Shanghai Institute of Applied Physics, Chinese Academy of Sciences, Shanghai, 201800}
\author{Hong-Wei Wang\orcidF{}}
 \affiliation{Shanghai Advanced Research Institute, Chinese Academy of Sciences, Shanghai, 201210}
 \affiliation{Shanghai Institute of Applied Physics, Chinese Academy of Sciences, Shanghai, 201800}
 \affiliation{University of Chinese Academy of Science, Beijing, 101408}

\author{Hang-Hua Xu\orcidG{}}
 \affiliation{Shanghai Advanced Research Institute, Chinese Academy of Sciences, Shanghai, 201210}
\author{Long-Xiang Liu\orcidH{}}
 \affiliation{Shanghai Advanced Research Institute, Chinese Academy of Sciences, Shanghai, 201210}
\author{Yue Zhang\orcidI{}}
 \affiliation{Shanghai Advanced Research Institute, Chinese Academy of Sciences, Shanghai, 201210}

\author{Jiunn-Wei Chen}
\affiliation{Department of Physics,
National Taiwan University, Taipei, 106}
\affiliation{InQubator for Quantum Simulation (IQuS), Department of Physics, University of Washington, Seattle, WA, USA, 98195}

\author{Yu-Xuan Yang\orcidK{}}
\affiliation{Shanghai Institute of Applied Physics, Chinese Academy of Sciences, Shanghai, 201800}
\author{Sheng Jin\orcidL{}}
\affiliation{Shanghai Institute of Applied Physics, Chinese Academy of Sciences, Shanghai, 201800}
\author{Kai-Jie Chen\orcidM{}}
\affiliation{Shanghai Institute of Applied Physics, Chinese Academy of Sciences, Shanghai, 201800}
\author{Zhen-Wei Wang\orcidN{}}
\affiliation{Shanghai Institute of Applied Physics, Chinese Academy of Sciences, Shanghai, 201800}
\author{Xiang-Fei Wang\orcidO{}}
\affiliation{Shanghai Institute of Applied Physics, Chinese Academy of Sciences, Shanghai, 201800}
\author{Meng-Ke Xu\orcidP{}}
\affiliation{Shanghai Institute of Applied Physics, Chinese Academy of Sciences, Shanghai, 201800}

\author{Zhi-Cai Li\orcidQ{}}
\affiliation{Shanghai Advanced Research Institute, Chinese Academy of Sciences, Shanghai, 201210}
\author{Pu Jiao\orcidR{}}
\affiliation{Shanghai Advanced Research Institute, Chinese Academy of Sciences, Shanghai, 201210}
\author{Meng-Die Zhou\orcidS{}}
\affiliation{Shanghai Advanced Research Institute, Chinese Academy of Sciences, Shanghai, 201210}
\author{Shan Ye\orcidT{}}
\affiliation{Shanghai Advanced Research Institute, Chinese Academy of Sciences, Shanghai, 201210}
\author{Yu-Long Shen\orcidU{}}
\affiliation{Shanghai Advanced Research Institute, Chinese Academy of Sciences, Shanghai, 201210}

\author{Yin-Ji Chen\orcidV{}}
\affiliation{Institute of Modern Physics, Fudan University, Shanghai, 200433}
\author{Hao Zhang\orcidW{}}
\affiliation{Institute of Modern Physics, Fudan University, Shanghai, 200433}
\author{Jian-Jun He\orcidX{}}
\affiliation{Institute of Modern Physics, Fudan University, Shanghai, 200433}



\author{Wen-Qing Shen}
 \affiliation{Shanghai Advanced Research Institute, Chinese Academy of Sciences, Shanghai, 201210}
 \affiliation{Shanghai Institute of Applied Physics, Chinese Academy of Sciences, Shanghai, 201800}

\author{Yu-Gang Ma\orcidB{}}
\email{mayugang@fudan.edu.cn}
\affiliation{Key Laboratory of Nuclear Physics and Ion-beam Application (MoE), Institute of Modern Physics, Fudan University, Shanghai 200433}
\affiliation{School of Physics, East China Normal University, Shanghai 200062}

\date{\today}

\begin{abstract}
The photodisintegration cross sections of the deuteron have been systematically measured over the photon energy range of 2.33--19.65 MeV at the Shanghai Laser Electron Gamma Source (SLEGS). By applying the well-established Baldin sum rule to the newly obtained data, the sum of the electric and magnetic dipole polarizabilities of the deuteron is extracted for the first time based solely on a dense and continuous experimental dataset, yielding $\alpha_E + \beta_M = 0.719 \pm 0.009_{\text{stat}} \pm 0.014_{\text{algo}} \pm 0.023_{\text{syst}}$ fm$^3$. With theoretical values of the magnetic polarizability $\beta_M$ calculated from the pionless effective field theory, a new value of the electric polarizability is obtained as $\alpha_E = 0.637 \pm 0.009_{\text{stat}} \pm 0.014_{\text{algo}} \pm 0.023_{\text{syst}} \pm 0.004_{\text{theo}}$ fm$^3$, which is in excellent agreement with current theoretical predictions. This result resolves the previous discrepancy between experimental measurements from elastic scattering and theory, providing a high-precision benchmark for nuclear interaction models.
\end{abstract}

\keywords{deuteron electric polarizability, deuteron photoneutron cross section, sum rule, SLEGS}
\maketitle



{\it Introduction.} The electric dipole polarizabilities ($\alpha_{E}$) of nucleons and light nuclei—such as the neutron~\cite{WOS:000348455000002,WOS:000182928300007,WOS:000174905400011,WOS:000085791900013,WOS:A1991EZ13400013,WOS:000087829800008,WOS:000350218000012}, proton~\cite{WOS:000308630000001,WOS:000371938900002,WOS:000804568000004}, and deuteron~\cite{WOS:000230889300013,Ji:2003ia}—characterize the response of their internal charge distributions to external electric fields~\cite{WOS:000226161000062}. The deuteron, composed of a proton and a neutron, is the simplest bound nuclear system and plays a fundamental role in constraining the nucleon-nucleon (NN) interaction potential~\cite{WOS:000827833400001}. Its photodisintegration cross sections serve as a key experimental tool for extracting parameters of the NN potential~\cite{arenhovel2012photodisintegration,WOS:A1977EE48300005,WOS:A1976CD52400011}. In particular, the electric dipole polarizability $\alpha_{E}$ dominates the integral that appears in the Baldin-type sum rules based on photodisintegration data~\cite{10.1007/978-3-7091-9453-9_37,Ji:2003ia}, thereby providing a crucial constraint on theoretical models of nuclear forces.

Experimentally, two principal approaches have been developed to determine $\alpha_{E}$ for the deuteron. The first involves elastic scattering of deuterons off heavy atomic targets, yielding an early measured value of $\alpha_E = 0.70 \pm 0.05$ fm$^3$~\cite{82exp}. However, this result is significantly larger than those predicted by theoretical calculations~\cite{PhysRev.128.2724,WOS:A1977EE48300005,WOS:A1983RM19700038,WOS:A1984RY55600025,WOS:A1990CU67000011,WOS:A1997XF03100003,Chen:1998vi,Ji:2003ia,WOS:000230889300013}. A similar overestimation has been observed for $^3$He, for which the measured polarizability using the same method is $0.25 \pm 0.04$ fm$^3$~\cite{WOS:A1991ET08600011}, nearly twice the theoretical expectation of 0.14 fm$^3$~\cite{WOS:000267701200010}. These consistent deviations suggest that the elastic scattering method may systematically overestimate $\alpha_E$, potentially due to complex atomic effects or model-dependent assumptions in interpreting scattering observables.

The second approach to determining $\alpha_E$ relies on applying a sum rule to experimental photon-induced deuteron disintegration cross sections, offering a more model-independent extraction. A notable early attempt in this direction was made in Ref.~\cite{83exp}. However, several important limitations undermine the reliability of its result. Most critically, the dataset used in Ref.~\cite{83exp} lacks photoneutron cross sections below $E_\gamma < 5$ MeV—a region that, as later confirmed, contributes significantly to the $\alpha_E$ integral. Furthermore, the method by which low-energy cross sections were estimated or extrapolated was not clearly explained. The study also adopted a uniform 6\% uncertainty for the photodisintegration data, despite these data originating from multiple experiments with differing uncertainties: 6\%, 7\%, and 14\% respectively~\cite{datasheet1983,WOS:A1973P329300011}. No rationale was given for selecting specific datasets or for homogenizing their uncertainties. Consequently, the extracted value of $\alpha_E$ in Ref.~\cite{83exp} is subject to questionable assumptions and likely underestimates its true uncertainty.

These challenges highlight the need for a new, high-precision measurement of the deuteron photodisintegration cross section with full low-energy coverage, which can be directly used in the sum rule analysis to yield a reliable and accurate value of $\alpha_E$.


Given the current status of the $\alpha_{E}$ measurements, we have conducted a re-evaluation of $\alpha_{E}$ from D($\gamma$,n)p photodisintegration through the sum rule method. In the D($\gamma$,n)p process, following the well-known Baldin sum rule \cite{WOS:A1960WQ60200010} and latest effective field theory \cite{Ji:2003ia}, the averaged photodisintegration cross section $\sigma_{-2}$ and electromagnetic polarizabilities can be written as 
\begin{equation}
\sigma_{-2}=\int^{\infty}_{0}\sigma(E_{\gamma})E_{\gamma}^{-2}\rm{d}\it{E_{\gamma}}=\alpha_{E}+\beta_{M} , \label{sigma_N2}
\end{equation}
where $\sigma(E_{\gamma})$ is the photodisintegration cross section. $E_{\gamma}$ is the incident photon energy. And $\beta_{M}$ is the magnetic polarizability of deuteron. In this paper, the D($\gamma$,n)p photodisintegration cross section was systematically measured for the first time in the entire range from the neutron emission threshold $S_\text{n}$ to 19.65 MeV through a precise energy scan, presenting the most densely populated and systematically measured dataset to date in the energy region below 20 MeV for the sum rule $\alpha_{E}+\beta_{M}$ calculation.

{\it Experimental set-up and methods.} The experiment was performed at the Shanghai Laser Electron Gamma Source (SLEGS) beamline \cite{WangHW,nsrpol}, which generates a MeV-scale gamma-ray beam via laser Compton slant-scattering (LCSS). In this process, photons from a 10.64 $\upmu$m CO$_2$ laser interact with the 3.5 GeV electron beam stored in the Shanghai Synchrotron Radiation Facility (SSRF) storage ring \cite{TaiRZ}. Benefiting from the SLEGS slant-scattering mode, a total of 83 energy points were densely sampled across a wide range from 2.33 MeV to 19.65 MeV. The CO$_2$ laser operates with a period of 1000 $\upmu$s and a pulse width of 50 $\upmu$s, resulting in a gamma-ray beam that retains the same temporal structure. The flux and energy profile of the SLEGS gamma-ray beam under various operating conditions have been characterized in previous studies \cite{XU2025170249,liu2024energy}. The incident gamma-ray flux and spectrum were monitored using a 76 mm $\times$ 200 mm BGO or a 3-inch $\times$ 4-inch LaBr$_3$ detector, which is placed at the downstream end of the beamline. A copper attenuator was installed upstream of the monitor to reduce the beam intensity, maintaining a count rate of approximately 1 kHz to match the calibration experiment conditions \cite{liu2024energy}. The incident gamma-ray energy spectrum for each measurement was subsequently obtained by applying the unfolding method detailed in Ref. \cite{liu2024energy}. The collimation of the gamma-ray beam was performed by the double-collimator system, in which the aperture of the second collimator defines its energy spread. The collimated LCSS gamma-rays, with a typical energy spread (FWHM) of 0.43 to 1.23 MeV, were then directed onto a cylindrical D$_2$O target (99.90\% purity) contained in an aluminum cell ($\Phi$10$\times$100 mm). 

For measurements where the incident gamma energy exceeded the neutron separation energy of $^{27}$Al (13.058 MeV), an identical container filled with deuterium-depleted water was used for background determination. (The separation energy of $^{16}$O is 15.664 MeV.) The emitted photoneutrons were detected by a Flat-Efficiency Detector (FED) array \cite{2025FED}. This array consists of 26 $^{3}$He proportional counters embedded in a polyethylene moderator. The counters are arranged in three concentric rings at distances of 65 mm (inner), 110 mm (middle), and 175 mm (outer) from the beam axis. The entire array, which has an average detection efficiency of approximately 42.10\% for the neutrons of interest, was precisely aligned along the beam axis. In the experiment, the widely accepted Ring-Ratio technique \cite{gheorghe2017photoneutron,utsunomiya2017direct,2019IAEA} was employed to determine the average detection efficiency of the FED for the photoneutrons.

\begin{figure}
\centering
\includegraphics[width=\linewidth]{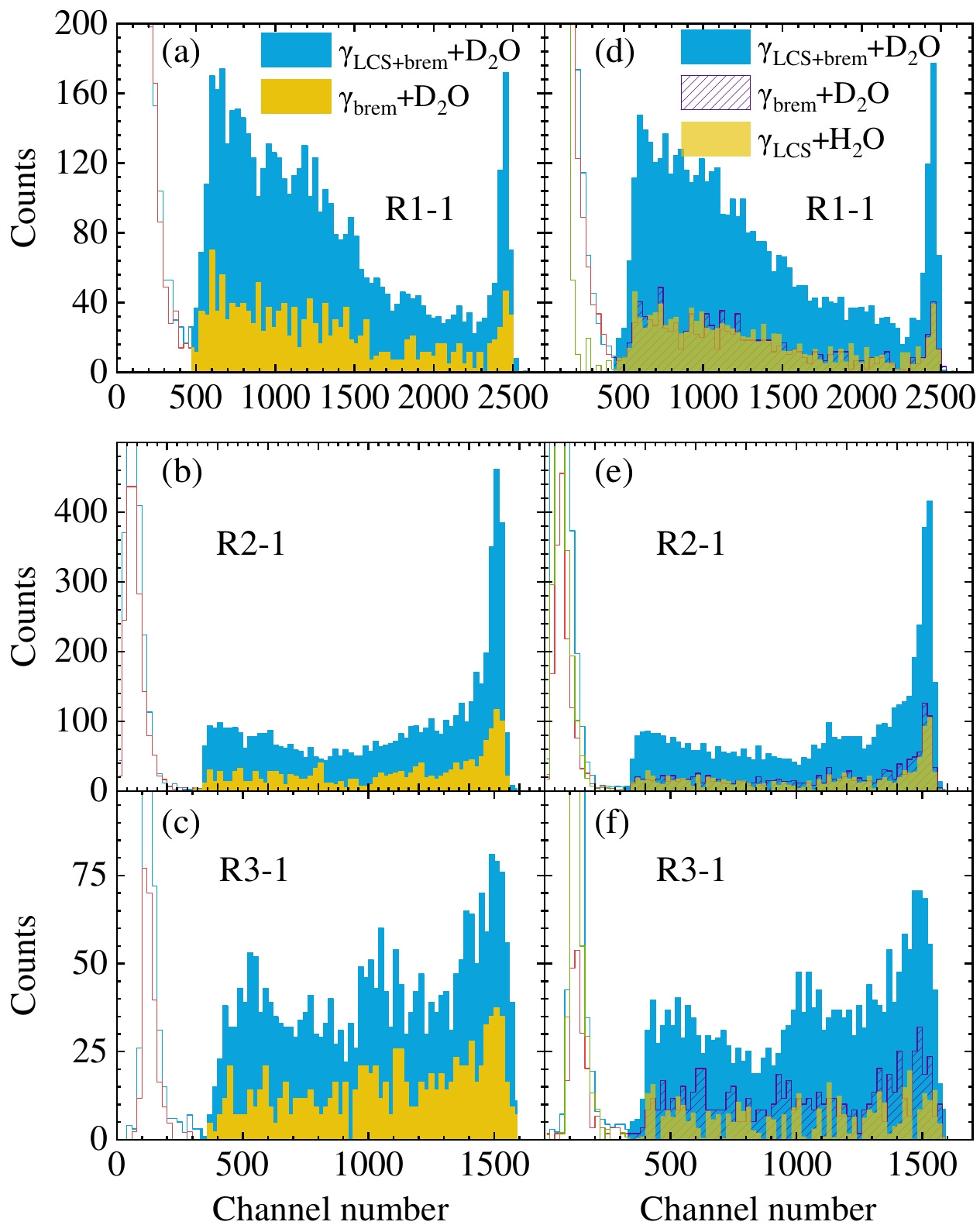}%
\caption{\label{He3spectra}
Typical pulse height spectra from $^{3}$He proportional counters. All spectra are normalized to a one-hour measurement time. The first column corresponds to a mean incident gamma energy of 5.584 MeV, and the second column to 16.127 MeV. Rows 1, 2, and 3 show the spectra from the first tube of the inner, middle, and outer rings, respectively. The unshaded peaks in the low‑channel region are the discriminated gamma‑ray background, separated from the neutron signals (shaded) by a pulse height threshold.  }
\end{figure}

The signals from the gamma-ray monitor detector were acquired using a CAEN DT5730B digitizer (14-bit, 500 MS/s), while those from the $^{3}$He proportional counters were recorded by a Mesytec MDPP16 digitizer. The pulsed nature of the gamma beam resulted in a distinct temporal structure for both the gamma and neutron signals: a prominent peak superimposed on a flat, time-independent background. This feature allowed for the subtraction of the non-pulsed background by selecting appropriate time gates in the time distribution spectra.

For the gamma-ray detector, the background-subtracted response spectrum was processed using the Direct Unfolding method to reconstruct the incident gamma-ray energy spectrum \cite{liu2024energy}. For the $^{3}$He proportional counters, the pulse height spectra contained contributions from both neutrons and gamma rays, with the latter dominating the lower pulse height region \cite{Knollbook}. A pulse height threshold was applied to discriminate neutron events from the gamma background, as illustrated in Fig. \ref{He3spectra}, where the neutron-induced signals are indicated by the filled colored regions. The shaded neutron regions in Fig. \ref{He3spectra} may contain a very small admixture of gamma rays, but this contamination is negligible and does not affect the final result. It should be noted that even for counters of identical specification, the applied threshold was individually determined for each tube due to slight variations in their amplification gain. These thresholds were determined through calibration measurements using a $^{252}$Cf neutron source, whose intense neutron yield relative to the in-beam conditions provided a clear separation between the neutron and gamma-ray components in the pulse height spectrum.

Fig. \ref{He3spectra} shows representative normalized pulse height spectra from single $^{3}$He proportional counter tubes in each of the three detector rings, all normalized to a one-hour measurement time. Column 1 corresponds to measurements where the incident gamma energy was below the neutron separation energy of $^{27}$Al. In this energy range, the background consists solely of neutrons generated from bremsstrahlung gamma-rays and can thus be isolated via time distribution analysis. The yellow-filled spectrum represents neutron events whose timing falls within the flat, constant-background region of the time distribution, corresponding to neutrons generated by bremsstrahlung gamma-rays. The blue-filled spectrum corresponds to neutron events within the time peak, representing neutrons from both LCS and bremsstrahlung gamma-rays. The net LCS neutron count detected by the FED is obtained by subtracting the background spectrum from the total spectrum and then integrating the resulting net spectrum. Column 2 corresponds to measurements where the incident gamma energy exceeded the neutron separation energy of $^{27}$Al (and, at higher energies, that of $^{16}$O). Under this condition, two distinct background sources contribute: neutrons produced by bremsstrahlung gamma-rays interacting with the D$_2$O, and neutrons generated by LCS gamma-rays interacting with nuclei in the target composition, primarily $^{27}$Al and $^{16}$O. The blue-filled spectrum represents the neutron response from the D$_2$O target due to both LCS and bremsstrahlung gamma-rays. The yellow-filled spectrum represents the background component from bremsstrahlung gamma-rays on the D$_2$O target. The spectrum shown as the shaded area represents another neutron background originating from LCS gamma-rays interacting with the H$_2$O target, which is used to account for contributions from the aluminum container and oxygen. This dedicated measurement with an H$_2$O target enables the complete subtraction of all LCS induced non-deuteron contributions, primarily from the aluminum container and oxygen in the target assembly. As shown in Fig. \ref{He3spectra}, the signal-to-noise ratio remains sufficient at each energy point despite the complex composition of the background, thereby enabling reliable subtraction of the background contributions.


\begin{figure}
\centering
\includegraphics[width=\linewidth]{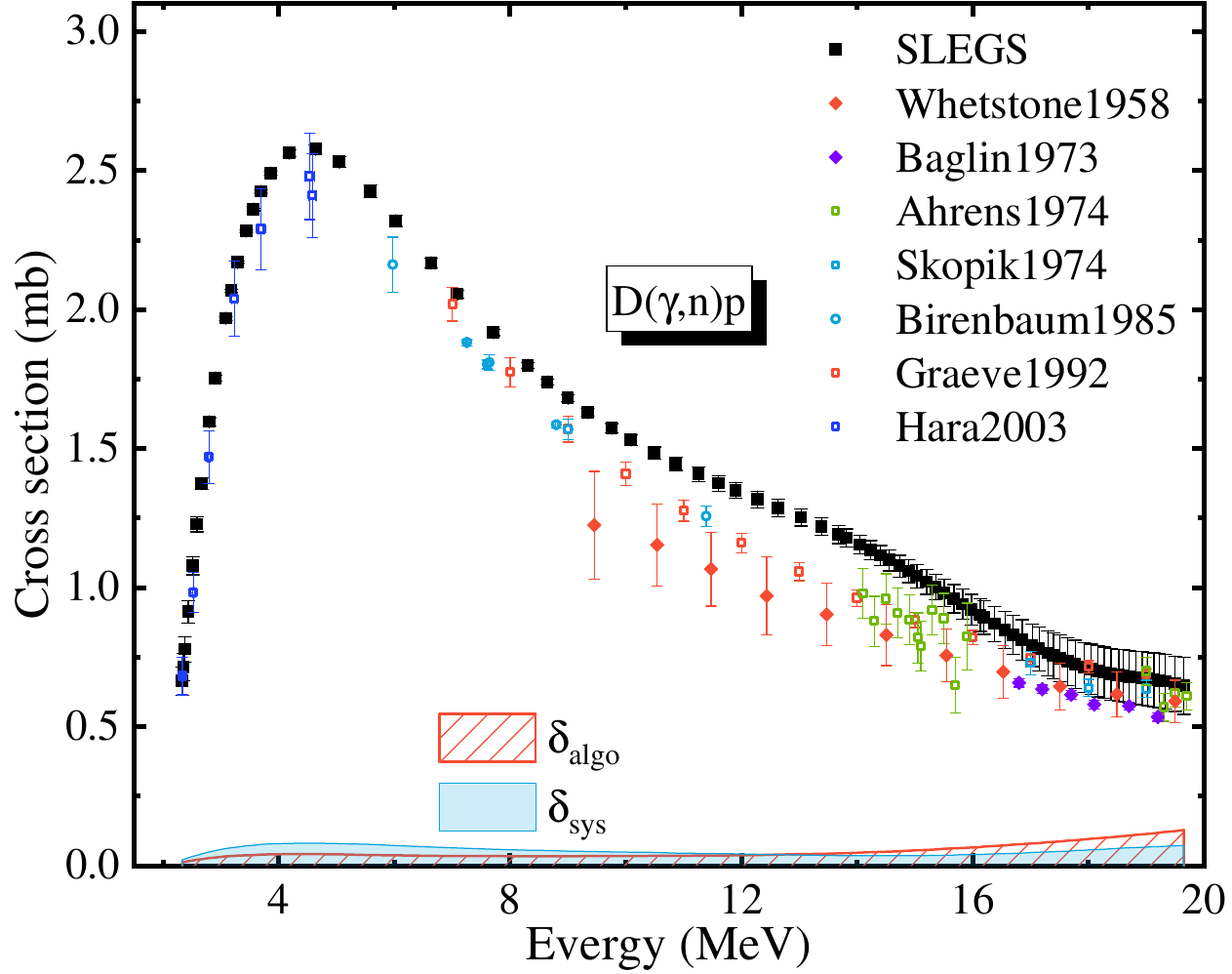}
\caption{\label{D_crosssection} The D($\gamma$,n)p photodisintegration cross sections measured from the present experiment. The error bars shown are the statistical uncertainties. The blue band is the systematic uncertainty. And the red band shows the algorithmic uncertainty.  }
\end{figure}


Following the preliminary data processing, an initial cross section can be calculated directly from its definition. Owing to the quasi-monoenergetic nature of the gamma-ray source, this is conventionally termed the quasi-monoenergetic cross section. In practice, this quantity represents a folded cross section, weighted over the incident gamma-ray energy spectrum. This folded D($\gamma$,n)p photodisintegration cross section, denoted $\sigma_{\rm{exp}}$, is defined as:


\begin{equation}
\sigma_{\rm{exp}}=\int^{E_{\text{max}}}_{S_\text{n}}n_{\gamma}(E_{\gamma})\sigma(E_\gamma)dE_\gamma = \frac{N_\text{n}}{N_\text{t}N_\gamma\xi\epsilon_\text{n}g}, \label{crosssectiondefination}
\end{equation}
\noindent
$n_{\gamma}(E_{\gamma})$ is the normalized energy spectrum incident on the target and the $\sigma(E_{\gamma})$ is the unfolded cross section. $N_\text{n}$ is the number of LCS neutrons detected by FED. $N_\text{t}$ is the number of the target nuclei per area. $N_\gamma$ is the number of gamma rays incident on the target. $\xi=(1-e^{\mu t})/(\mu t)$ is a correction factor for thick target. Here, $\mu$ denotes the folded mass attenuation coefficient, which is derived by combining the mass attenuation coefficients listed in \cite{NISTweb} with the incident gamma-ray spectrum. $t$ is the target thickness. $\epsilon_\text{n}$ is the neutron detection efficiency of the FED array, determined using the Ring-Ratio technique \cite{gheorghe2017photoneutron,ITOHAuCrossSection,krzysiek2019photoneutron,utsunomiya2017direct,2019IAEA}. The reliability of this efficiency has been verified through independent calibration measurements using a $^{252}$Cf neutron source. $g$ represents the proportion of the energy spectrum above the neutron threshold.


{\it Results and discussion.} 
To accurately determine the electric polarizability $\alpha_{E}$ via the sum rule, high-precision photodisintegration cross section data are essential, which in turn requires stringent control over systematic uncertainties. In this work, the main systematic uncertainties originate from the FED detection efficiency, Cu attenuator, and the target thickness. The individual contributions are quantified as follows:
(1) FED Detector Efficiency: The total efficiency uncertainty of the FED array is 3.02\%. This value results from a careful quadrature sum of contributions from: the emission rate of the $^{252}$Cf calibration source (3.00\%), high-voltage supply fluctuations (0.02\%), target positioning shifts (0.10\%), data acquisition parameter settings (0.17\%), and system stability (0.26\%).
(2) Cu Attenuator Thickness: The uncertainty in the effective thickness of the Cu attenuator impacts the reconstructed incident gamma-ray flux. While the inherent thickness tolerance is negligible, potential alignment deviations when positioning the attenuator and considerations of material purity contribute to a combined uncertainty of less than 1.00\% in the effective attenuation.
(3) D$_2$O Target: The uncertainty associated with the physical length of the target cell is negligible. The dominant contribution arises from the stated purity of the D$_2$O, which is 0.10\%.
(4) Gamma-ray Flux Monitoring: The systematic effect from the BGO monitor detector is negligible due to its stable high-voltage supply, well-characterized calibration, and near-100\% detection efficiency for the gamma-ray energies of interest. In contrast, the LaBr$_3$ detector does not have 100\% efficiency. A comparative analysis showed that the reconstructed incident flux using the LaBr$_3$ detector differs by approximately 2\% from that obtained with the BGO detector \cite{liu2025energy}. Therefore, a 2\% systematic uncertainty is assigned when the LaBr$_3$ detector was used.

In addition to the systematic uncertainties, other sources were evaluated: 1. Statistical Uncertainty. The statistical uncertainty is dominated by the finite neutron counts from the FED array, while that from the gamma-ray monitor is negligible due to its high count rate. 2. Algorithmic Uncertainty. This arises from data processing procedures. In extracting $N_\text{n}$, the choice of the background time region introduces an uncertainty, which is quantified to be 2.00\%. In determining $N_{\gamma}$, the unfolding process depends on a simulated detector response matrix, leading to an estimated uncertainty of 1.00\%.

The folded cross section $\sigma_{\rm{exp}}$ in Eq. \ref{crosssectiondefination} serves as the input reference value for extracting the monoenergetic cross section via the unfolding iteration method \cite{renstrom2018verification}. Its derivation depends on the incident gamma-ray energy relative to the neutron separation energy of $^{27}$Al. For energies below this threshold, the folded monoenergetic cross section can be obtained directly according to Eq. \ref{crosssectiondefination}. At higher energies, where non-deuteron backgrounds become significant, the folded cross section is instead calculated via the subtraction $\sigma_{\rm{exp}} = (\sigma_{\rm{exp}}^{\rm{D_2O}} - \sigma_{\rm{exp}}^{\rm{H_2O}})/2$, where $\sigma_{\rm{exp}}^{\rm{D_2O}}$ and $\sigma_{\rm{exp}}^{\rm{H_2O}}$ denote the folded cross sections for the D$_2$O and H$_2$O targets, respectively. This subtraction is valid provided the incident gamma-ray flux is consistent for both target measurements. In our experiment, this condition was ensured by sequentially measuring the D$_2$O and H$_2$O targets with all other experimental conditions unchanged.

\begin{figure}
\centering
\includegraphics[width=\linewidth]{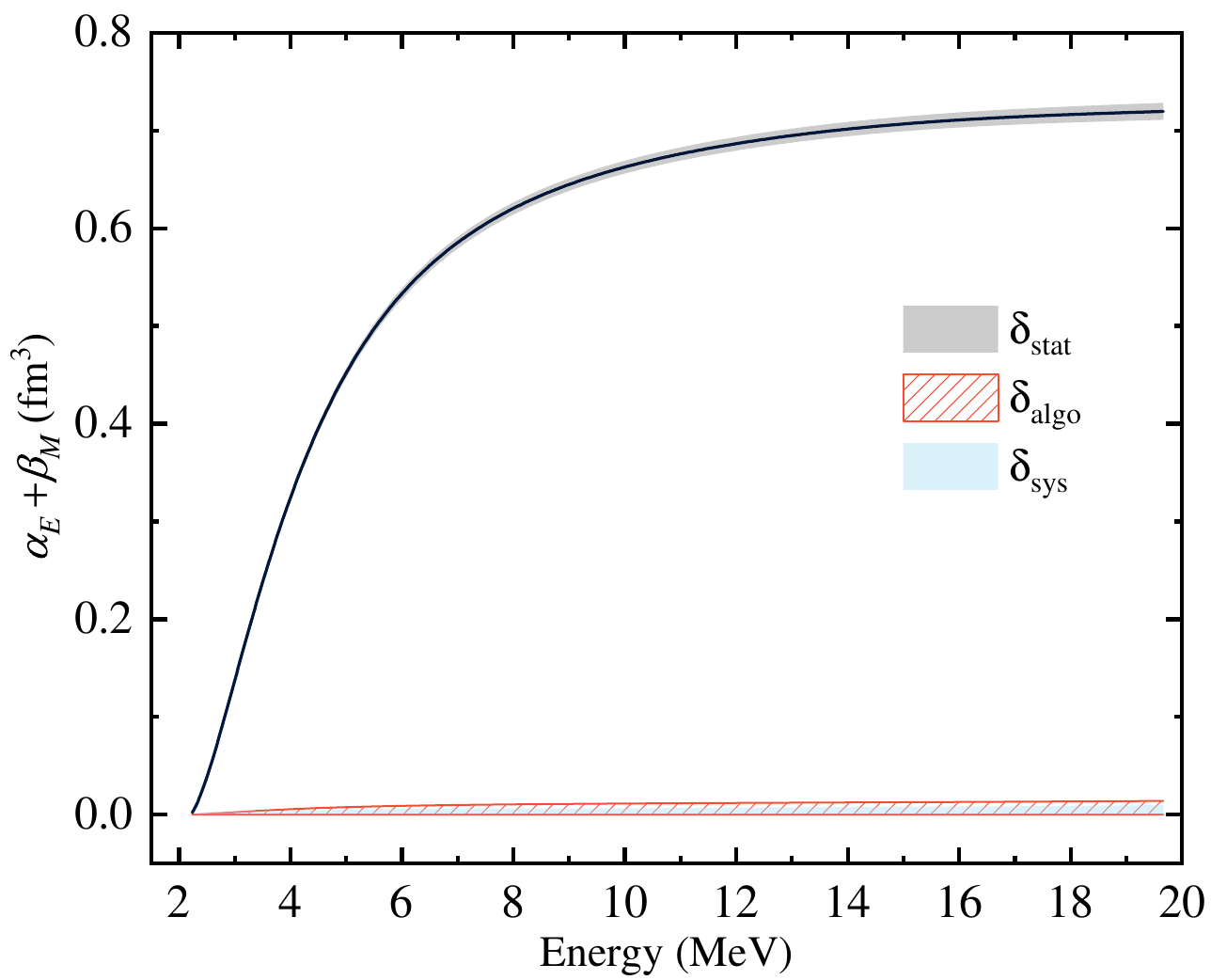}
\caption{\label{sigma_N2_E} The $\alpha_{E}+\beta_{M}$ as the function of the integral upper limit energy. The curve with gray band are the statistical uncertainties. The blue band is the systematic uncertainty. And the red band shows the algorithmic uncertainty. Note the sharp rise below 6 MeV, which emphasizes the crucial role of low-energy data in sum rule evaluations.}
\end{figure}

The unfolded cross section was subsequently derived using an established unfolding iteration procedure, detailed in previous works \cite{AuTbScibull, WOS:001397928300001, 59Coprc}. The final unfolded D($\gamma$,n)p cross section is presented in Fig. \ref{D_crosssection}. The total uncertainty is visually decomposed: the blue band represents the systematic uncertainty, the red shading indicates the algorithmic uncertainty, and the error bars on the black data points reflect the statistical uncertainty alone. The notable increase in total uncertainty near 18 MeV is attributed to the decreasing D($\gamma$,n)p cross section coupled with the growing background contribution from the $^{16}$O($\gamma$,n) reaction.

From Fig. \ref{D_crosssection}, we provide for the first time a complete and continuous coverage of the  D($\gamma$,n)p cross section measurement across the low-energy region, addressing the previous lack of systematically covered data in this range. 
The results are in overall agreement with the existing discrete data sets from various earlier studies in this energy domain. 
Regarding the specific energy interval of 10--15 MeV, the present results are higher (by approximately 2$\sigma$) than the earlier measurements of Skopik (1974) \cite{WOS:A1974S132600010}, Birenbaum (1985) \cite{PhysRevC.32.1825}, and Graeve (1992) \cite{WOS:A1992HD25700044}, while those three data sets are mutually consistent. It should be noted that outside this interval, i.e., at 6--10 MeV and 15--20 MeV, our data agree with the aforementioned measurements as well as with other existing data within the quoted uncertainties. Furthermore, the $^{197}$Au($\gamma$,n) cross section for the standard reaction obtained simultaneously in this experiment shows agreement with established data sets, confirming the reliability of our experimental setup and analysis procedure. A more comprehensive comparison and future measurements could help clarify the experimental situation in this energy range.

Then the D($\gamma$,n)p photodisintegration cross section obtained in this work were used to extracted the $\sigma_{-2}$, namely $\alpha_{E}+\beta_{M}$, through the well-known Baldin sum rule according to Eq. \ref{sigma_N2}. Fig. \ref{sigma_N2_E} shows the cumulative value of $\alpha_{E}+\beta_{M}$ as the function of the upper integration limit. The sum increases rapidly for upper limits below 6 MeV, underscoring that precise cross section measurements in this low-energy region are fundamental for an accurate determination. Although the $^{16}$O($\gamma$,n) reaction threshold near 16 MeV introduces an additional source of background, the integration beyond this energy still contributes significantly to the final sum, as is evident from Fig. \ref{sigma_N2_E}. Therefore, this region must be included. It is worth to note that due to the $E_{\gamma}^{-2}$ weighting factor in the integration, the contribution from the cross section diminishes rapidly with increasing energy. Consequently, the impact of the unmeasured high-energy region (from 19.65 MeV to infinity) is inherently suppressed. This contribution was subsequently estimated using existing experimental data from the EXFOR database to ensure the completeness of the integral. Specifically, a fit to these data was performed and used to evaluate the high-energy contribution to $\sigma_{-2}$ via Eq. \ref{sigma_N2}. The integral over the range 19.65--780 MeV yields a contribution of 0.012 fm$^{3}$, which much smaller than the total experimental uncertainty. The final result of the $\sigma_{-2}$ is:

\begin{equation}
\begin{split}
\alpha_{E}+\beta_{M}=0.719\pm0.009_{\rm{stat}}\pm0.014_{\rm{algo}}\pm0.023_{\rm{syst}}\   \text{fm}^{3}.
\end{split}
\end{equation}

\begin{figure}
\centering
\includegraphics[width=\linewidth]{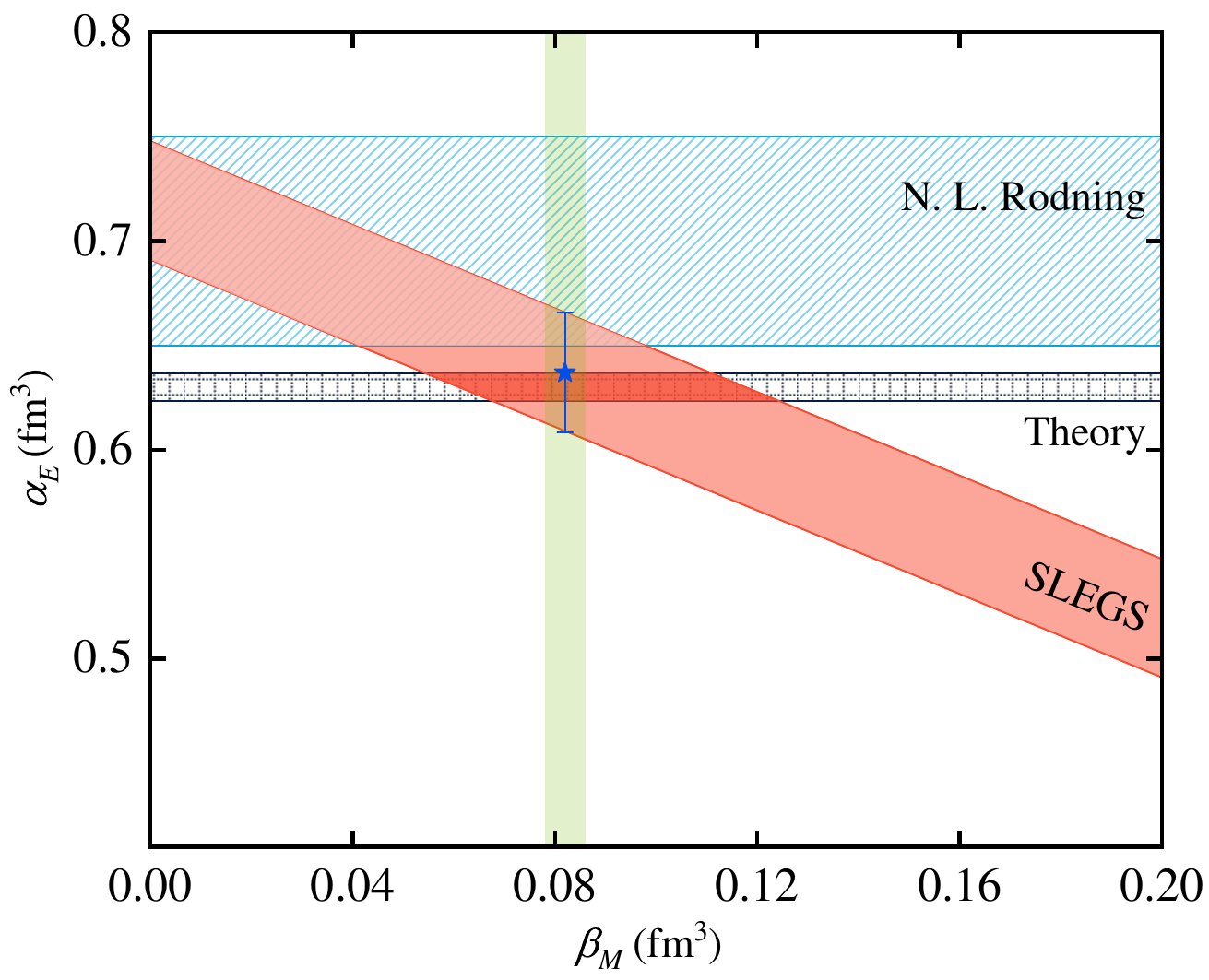}
\caption{\label{Compare_sigma_N2_E} 
The electric polarizability of the deuteron. The red band represents the $\alpha_{E}+\beta_{M}$ extracted from SLEGS D($\gamma$,n)p cross sections. The blue band shows the result from N.L. Rodning \cite{82exp}. The black grid represents the results of calculations from various theoretical models, with the error band corresponding to twice the standard deviation. The vertical green band denotes the theoretical calculations for $\beta_{M}$ and its associated uncertainty. The blue dot with an error bar indicates the $\alpha_{E}$ of this work.
}
\end{figure}

The results of the present work are compared with previous experimental and theoretical determinations in Fig. \ref{Compare_sigma_N2_E}. A number of theoretical calculations of the deuteron electric dipole polarizability, $\alpha_E$, have been reported, including values of 0.63 fm$^3$ \cite{PhysRev.128.2724}, 0.615 fm$^3$ 0.622 fm$^3$ and 0.628 fm$^3$\cite{WOS:A1977EE48300005}, 0.639(6) fm$^3$ \cite{WOS:A1983RM19700038}, 0.632(3) fm$^3$ \cite{WOS:A1984RY55600025}, 0.627 fm$^3$ \cite{WOS:A1990CU67000011}, 0.6328(17) fm$^3$ and 0.6334(14) fm$^3$ \cite{WOS:A1997XF03100003}, 0.6339 fm$^3$ \cite{Ji:2003ia}, 0.6378(12) fm$^3$ \cite{WOS:000230889300013}. To provide a consolidated theoretical reference, the mean of these values is calculated to be 0.630 fm$^3$, with a standard deviation of 0.007 fm$^3$ (denoted $\Delta\alpha_E^{\rm{theo}}$), and is represented by the black hatched band in the figure.

The blue band shows the result from elastic scattering experiments \cite{82exp}. A clear discrepancy exists between the elastic scattering result and the theoretical predictions, quantified as a difference of 1.386 $\sqrt{(\Delta\alpha_E^{\rm{exp}})^2 + (\Delta\alpha_E^{\rm{theo}})^2}$.


The red band indicates our present result for the sum $\alpha_{E}+\beta_{M}$, with its total uncertainty (statistical, algorithmic, and systematic). To extract $\alpha_E$ from this sum, the magnetic polarizability $\beta_M$ is required. In this study, the $\beta_M=0.082\pm0.004$ fm$^3$ is determined from the pionless effective field theory \cite{Phillips:1999hh,Chen:1998vi,Chen:1999tn,Chen:1999bg,Rupak:1999rk}, as detailed in the supplemental material \cite{supplementalmaterial}. The vertical green band in Fig. \ref{Compare_sigma_N2_E} indicates the theoretical $\beta_M$ and its uncertainty. The overlap region between the red and green regions represents the final result for $\alpha_{E} $,

\begin{equation}
\begin{split}
\alpha_{E}=0.637&\pm0.009_{\rm{stat}}\pm0.014_{\rm{algo}}\\
&\pm0.023_{\rm{syst}}\pm0.004_{\rm{theo}}\ \text{fm}^{3}.
\label{Pol}
\end{split}
\end{equation}

Our result for $\alpha_E$ is in good agreement with the theoretical prediction. The difference between the central values is quantified as 0.237 $\sqrt{ (\Delta\alpha_E^{\rm{exp}})^2 + (\Delta\alpha_E^{\rm{theo}})^2 }$, indicating statistical consistency.

{\it Conclusion.} 
In summary, we have performed a new systematic and high-precision measurement of the D($\gamma$,n)p photodisintegration cross section at the Shanghai Laser Electron Gamma Source, covering the photon energy range from 2.33 to 19.65 MeV with unprecedented resolution and density. For the first time, the sum of the electric and magnetic polarizabilities of the deuteron, $\alpha_{E} + \beta_{M}$, was directly extracted from experimental data using the Baldin sum rule.
By combining our experimental result with the theoretical value of $\beta_{M}$ from pionless effective field theory, we have obtained a new experimental value for the electric dipole polarizability $\alpha_{E}$. The extracted value shows excellent agreement with state-of-the-art theoretical calculations, effectively resolving the long-standing discrepancy between theory and earlier experimental results derived from elastic scattering of deuterons off heavy nuclei.


{\it Acknowledgments.} 
We would like to thank professor Hiroaki Utsunomiya and professor Xiangdong Ji for their insightful comments and constructive suggestions. This work was supported by the National key R\&D program (No.2023YFA1606901), the National Natural Science Foundation of China (No.12388102, No.12275338, No. 12547102).



\bibliography{apssamp}

\end{document}